\documentclass[aps,pre,reprint,twocolumns,notitlepage,superscriptaddress]{revtex4-1}
\usepackage{amssymb,bm}
\usepackage{amsmath,eucal}
\usepackage[usenames,dvipsnames]{color}
\usepackage{graphicx}
\usepackage{units}

\newcommand*{\VEC}[1]{\boldsymbol{#1}}
\newcommand*{\TENSOR}[1]{\mathsf{#1}}
\newcommand*{\OP}[1]{{\cal{#1}}}

\newcommand*{\EPS}{\varepsilon}

\begin{document}

\title{Steady diffusion in a drift field: a comparison of 
large deviation techniques and multiple-scale analysis}
\author{Erik Aurell}
\affiliation{
KTH -- Royal Institute of Technology,  AlbaNova University Center, SE-106 91~Stockholm, Sweden
}
\affiliation{Depts. Information and Computer Science and Applied Physics, Aalto University, Espoo, Finland}
\author{Stefano Bo}
\affiliation{Nordita, Royal Institute of Technology and Stockholm University,
Roslagstullsbacken 23, SE-106 91 Stockholm, Sweden}
\begin{abstract}
A particle with internal unobserved states
diffusing in a force field will generally display
effective advection-diffusion. The drift velocity
is proportional to the mobility averaged over the internal
states, or effective mobility, while the effective diffusion has
two terms. One is of the equilibrium type and
satisfies an Einstein relation with the
effective mobility while the other 
is quadratic in the applied force.
In this contribution we present two new methods to obtain these
results, on the one hand using
large deviation techniques, and on the other by
a multiple-scale analysis, and compare the two.
We consider both systems with discrete internal states and
continuous internal states.
We show that the auxiliary equations in the
 multiple-scale analysis can also be derived in second-order
perturbation theory in a large deviation theory of a generating function
(discrete internal states)
or generating functional (continuous internal states). 
We discuss that measuring the two components of the effective diffusion
give a way to determine kinetic rates from only first and second moments
of the displacement in steady state. 
\end{abstract}
\date{\today}
\maketitle
\section{Introduction}
Measuring on microscopic and mesoscopic scales is the art of rendering visible what we cannot directly see.
If the phenomenon of interest is faster than your time resolution you will only
see a blur, but what happens on the faster scale may still be crucial.
For instance, the fraction of the time a protein can be found in a given state
is proportional to the Gibbs factor, which is generally
the ratio of an on rate and an off rate, 
but the speed of catalyzed reactions
depend on the kinetic rates directly~\cite{english2006}. 
Similarly, the average occupancy of  transcription factors to their binding sites
on DNA determines the level of gene expression,
but the time scales of regulation of transcription are
set by the on rates and off rates separately~\cite{Buchler2003,Alon}. 
State-of-the-art methods to measure kinetic rates
rely on precise imaging of the molecules motion and translate
dynamic changes of their diffusion into transition
rates (see \cite{persson2013,persson2013review,hammar2014} for {\it in vivo} measurements and \cite{terazima2006} for an {\it in vitro} method). 
These methods have had great impact in determining the actual time scales and mechanisms of gene regulation
and elucidating classical issues in quantitative molecular biology such as how a transcription factor finds its
binding sites on DNA~\cite{BergvonHippel85,BergvonHippel89,HalfordMarko2004,hammar2014}.
They are however challenging to use, and other approaches, even if of more
limited utility, may also be of value. 

In this paper we study a non-equilibrium
effect, the dependence of the effective diffusion on an external applied force.
The effect can hardly be said to be new. In fact, to our best knowledge
it was first 
demonstrated by K.J.~Mysels in a paper published a little more than 60 years ago~\cite{Mysels56}.
This author observed that if particle or system transits
between two internal states as $
+\rightleftharpoons 
-
$ with rates
$k_+$ and $k_-$, and if the mobilities of the particle or system in the
two states are $\mu_+$ and $\mu_-$, then in a force field $F$ the effective diffusion is
\begin{equation}
\label{eq:Mysels}
D^{eff} = \frac{ T\left(k_+ \mu_- + k_- \mu_+\right)}{k_++k_-} + \frac{ F^2\left(\mu_+-\mu_-\right)^2 k_+k_-}{(k_++k_-)^3}
\end{equation}
In the first term $T$ is the temperature in units such that Boltzmann's constant $k_B$ is equal to unity.
The second term, in~\cite{Mysels56} called the ``electrodiffusion coefficient'', 
is the a priori surprising term, which cannot simply be postulated by analogy
to fluctuation-dissipation phenomena close to equilibrium.
Similar terms were discussed quite some time ago by Van den Broeck and co-workers
and by Mackey and co-workers, see~\text{e.g.}~\cite{VanDenBroeck83,VanDenBroeck84,Mackey97},
and recently reconsidered in the context of Taylor dispersion~\cite{taylor1953,vandenbroeck1990,kahlen2017}.
The effective diffusion of a protein was recently analyzed in \cite{illien2016diffusion} and applied to fluorescence correlation spectroscopy (FCS) measurements in \cite{illien2017exothermicity}.
These investigations concern non-equilibrium phenomena where the protein functions as an enzyme acting on a substrate but they do not feature any externally applied force.

The first goal of the present paper is to present
(\ref{eq:Mysels}) from a modern perspective.
We will derive it both by 
means of the scaled cumulant generating function 
and 
by a multiple-scale analysis. 
The cumulant 
technique also gives access to higher moments
of the displacement.
We will also show that a similar behavior as (\ref{eq:Mysels}),
which we recently found for the case of a rotating Brownian
particle with a continuous internal state (the particle's orientation)~\cite{aurell2016}
can also be derived from a large deviation argument and 
second order time-independent perturbation theory.
The second goal of the paper is to bring effects such as
(\ref{eq:Mysels}) anew to the attention of the single-molecule
biophysics community, as a possible means to indirectly
measure kinetic rates. Although established a long time ago
these non-equilibrium steady-state relations appear to have been
somewhat occluded, and the recent revival of interest has been 
motivated by other questions.  The analysis of Taylor dispersion in \cite{kahlen2017} is for instance partly foreshadowed by an analysis of active Brownian motion (Janus particles) \cite{pietzonka2016extreme} using cumulant generating functions, from which expressions like Eq.~\eqref{eq:Mysels} can be derived.

The paper is organized as follows.
In Section~\ref{sec:model} we define the model
and in Section~\ref{sec:heuristic} we 
give a heuristic analysis
that establishes a term like the second 
one in (\ref{eq:Mysels}), with a quadratic dependence
on the applied force.
In Section~\ref{sec:LD} 
we derive this result systematically using
large deviation theory \cite{varadhan2008large,touchette2009large}, and 
in Section~\ref{sec:multiple-scales} we obtain the results by means of multiple-scales analysis \cite{pavliotis08}.
In Section~\ref{sec:generalizations} we generalize to
space-dependent switching rates and to several internal states, details 
of the latter analysis given in Appendix~\ref{appendix:several}.
In Section~\ref{sec:device} we discuss the effect
as a device to measure kinetic rates and in Section~\ref{sec:conclusions-discussion}
we sum up and discuss our results.

Some technical material
for the explicit solutions of section ~\ref{sec:multiple-scales}
is given in Appendix~\ref{sec:eigen}. 

In two additional Appendices~\ref{appendix:translations} and~\ref{appendix:higher-moments}
we extend for completeness the large deviation analysis to the case of
a continuum of internal states, of which one
example is a Brownian particle undergoing translation and rotation.
We show that an auxiliary equation which appeared in an earlier multiple-scale analysis~\cite{aurell2016,marino2016b}, 
can also be derived by second-order perturbation theory of generating function of the time
spent in different orientations.
We further show that the large deviation analysis can also be pushed to the third centered moment
which we show to generically increase linearly in time, see Appendix~\ref{appendix:higher-moments}.

\section{Model}
\label{sec:model}
\begin{figure}[h]
\includegraphics[width=\columnwidth]{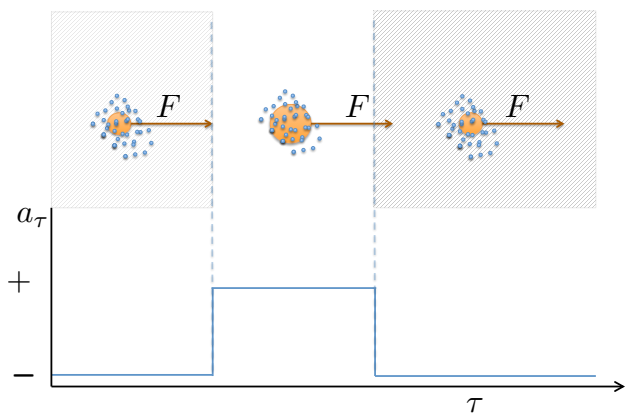}
\caption{%
Sketch of the system under consideration. A particle undergoes diffusion subject to a constant force. The friction of the particle changes stochastically in time switching between two states.}
\label{fig:scheme}      
\end{figure}
To illustrate our method we consider a particle diffusing in a fluid at temperature $T$ subject to an external constant force $F$ (Boltzmann constant set to 1).
The particle can switch between two different states ($-$ and $+$) associated with different mobilities ($\mu_-$ and $\mu_+$) as sketched in Figure~\ref{fig:scheme}.
The system can be described by a Langevin equation
\begin{equation}\label{eq:lang}
d x_t=F\mu_{a}dt+\sqrt{2T\mu_{a}}dW_t
\end{equation}
where the internal variable $a$ can be in either of two states denoted $-$ and $+$ and $dW_t$ is a Wiener increment.
The internal state changes Markovianly
with exponentially distributed waiting times and with transition rates
$k_+$ (from state $+$ to $-$) and $k_-$ (for the reversed transition $-$ to $+$) that are independent of spatial coordinate $x$.
For simplicity we consider constant rates in time. 
 The probability of finding the internal variable in a given state then evolves according to the master equation:
\begin{equation}\label{eq:master}
\frac{d}{dt}
\left(
 \begin{array}{c }
P_{+}	 \\
 P_{-}
\end{array}\right)
=\left( \begin{array}{c c }
-k_+& k_-	 \\
k_+& -k_-	
\end{array}
\right)
\left( \begin{array}{c }
P_{+}	 \\
 P_{-}
\end{array}
\right)=\TENSOR{K}\left( \begin{array}{c }
P_{+}	 \\
 P_{-}
\end{array}\right)
\end{equation}
and the steady state probability reads:
\begin{equation}\label{eq:ss}
w_+=\frac{k_-}{k_-+k_+}\qquad 
w_-=\frac{k_+}{k_-+k_+}\,.
\end{equation}
The joint probability density of being in position $x$ and state $a$ is determined 
by mixed diffusion and Master equation:
\begin{eqnarray}\label{eq:FP}
\frac{\partial  P_{a}(x)}{\partial  t}&=&
\frac{\partial}{\partial x}\left[\left( -\mu_{a}F+T\mu_{a}\frac{\partial}{\partial x} \right)P_{a}(x)\right]\\\nonumber
&+&k_{-a}P_{-a}(x)-k_aP_{a}(x)\;.
\end{eqnarray}
Our aim is to describe the statistics of the particle displacement on time scales much longer than the typical
relaxation of the discrete internal state.

\section{Heuristic analysis}
\label{sec:heuristic}
The gist of the idea can be explained in simple terms as follows. 
At the microscale, the mobility of a body determines how its motion responds to an applied
 force
and, together with temperature sets the magnitude of the thermal fluctuations, see~\eqref{eq:lang} above.
 If the particle dynamically switches between states with different mobilities, how far the body moves under a force 
  will depend on the time spent in each state.
The average displacement is determined by the magnitude of the external force and by the
effective mobility which is set by the mobility in each internal state weighted by the average fraction of time spent in a state. 

On top of such averages are fluctuations in how far the body travels.
Part of these fluctuations are thermal, due to the collisions with the fluid molecules in which the body is diffusing, and
therefore proportional to the effective mobility (Einstein relation).
However, given the presence of the external force, there are also fluctuations
caused by the switching between states of different mobility. 
Indeed, a particle that has been more (less) often than usual in the state with the larger friction will lag behind (be ahead of) the average position, and this increases the spread around the average position in a way that is  analogous to the phenomenon known as Taylor dispersion \cite{taylor1953,vandenbroeck1990,kahlen2017}.
Such contribution is  proportional to the variance of the time spent in each state and is 
 quadratic in the applied force.
%
In the rest of this paper we will discuss this issue more quantitatively, determine the prefactor
of the square of the force that depends on the kinetic rates, and compare and contrast
two ways of carrying out the analysis.
To illustrate the problem we will start with the simple case in which the internal variable can take only two states and
then generalize it.

\section{Large deviation analysis}
\label{sec:LD}
In this section we will use the scaled cumulant generating function to turn the above qualitative argument
into a quantitative prediction.
The long term statistics of the particle displacement can be computed by considering the explicit expression of its probability.
The dynamics of the discrete internal variable is not influenced by the position of the particle.
Its trajectories from an initial time $0$ to a final one $t$ are specified by the sequence of states
$
a(\tau)$ with   $0 \le \tau \le t$.
 Let us number the interval between transitions as $n$ and the transition times as  
 $\tau_n$. We then have that 
 in an interval $n$, of duration $\tau_{n+1}-\tau_n$, the system resides in state $a_n$ 
\begin{eqnarray*}
\left\{  a_0  \mbox{ in } (0=\tau_0,\tau_1) ; \cdots;   a_n \mbox{ in } (\tau_{n},\tau_{n+1}); \cdots;\right.
\\
\left. \cdots;a_{N}  \mbox{ in } (\tau_{N},\tau_{N+1}=t)  \right\}
\end{eqnarray*}
as depicted in Figure \ref{fig:scheme}.
Between transitions, the internal variable is fixed at its value $a_n$ and the particle undergoes a Langevin dynamics with a fixed value
of the mobility $\mu_{a_n}$.
The probability density of observing a displacement $\Delta x_{n}$ in the time interval $\Delta \tau_n=\tau_{n+1}-\tau_n$ in which $a=a_n$ is a simple Gaussian:
\begin{equation}\label{eq:p(x|an)}
P(\Delta x_n|a_n; \Delta \tau_n)=\frac{1}{\sqrt{4\pi \Delta \tau_nT\mu_{a_n}}}e^{-\frac{\left(\Delta x_n-F\mu_{a_n} \Delta \tau_n\right)^2}{4\Delta \tau_nT\mu_{a_n} }}\,.
\end{equation}
The spatial increments $\Delta x_n$ depend on the realizations of the Wiener noise
in the corresponding time intervals $\Delta\tau_n$, and
are therefore independent random variables.
The probability of overall displacement  $\Delta x=\sum_n \Delta x_n$ is hence also Gaussian
\begin{equation}\label{eq:p(x|a)}
P(\Delta x|\{a\}; t)=\frac{1}{\sqrt{4\pi T\sum_n\Delta \tau_n\mu_{a_n} }}e^{-\frac{\left(\Delta x-F\sum_n\mu_{a_n} \Delta \tau_n\right)^2}{4T\sum_n\Delta \tau_n\mu_{a_n} }}\,.
\end{equation}
The only information about the dynamics of the discrete internal variable affecting the displacement probability is the total time spent in 
one of the two states, {\it e. g.}, $\tau_+=\sum_n\tau_n\delta_{a_n}^+$  where $\delta_{a_n}^+$ is an indicator function that is $1$ when the discrete state
 $a_n=+$ and $0$ otherwise.
Introducing the the fraction of time spent in the $+$ state: $f$
 and recalling that $\sum_n\tau_n=t$, expression \eqref{eq:p(x|a)} simplifies to
\begin{eqnarray}\label{eq:p(x|f)}
P(\Delta x|\{a\};t)=P(\Delta x|f;t)
=\frac{
e^{-\frac{\left(\Delta x-\langle \Delta x|f\rangle\right)^2}{4tD(f)}}}{\sqrt{4 \pi tD(f)}}
\end{eqnarray}
where we have introduced the average particle displacement conditioned on spending a fraction $f$ of the total time $t$ in state $+$ 
\begin{equation}\label{eq:def_x}
 \langle \Delta x|f\rangle =Ft\mu(f)
 \end{equation}
 and
 \begin{equation}\label{eq:def_d}
 D(f)\equiv T\mu(f)
= \frac{T}{tF}\langle \Delta x|f\rangle
\end{equation}
with
\begin{equation}
\mu(f)=f \mu_{+}+(1-f)\mu_{-}
\end{equation}
which is linear in $f$.
In principle, it is then possible
 to compute the marginal probability of the body displacement 
by integrating the conditional expression \eqref{eq:p(x|f)}  multiplied by
 $P(f)$, the probability of the fraction of time spent in the $+$ state, also known as the
empiric (sample) average of state $+$:
\begin{eqnarray}\label{eq:p_marg}
P(\Delta x)=
\int P(\Delta x|f)P(f) df\,.
\end{eqnarray}
Despite the apparent simplicity of the problem, the explicit expression for  $P(f)$ 
 is rather complicated (see e.g. \cite{giddings55}) and 
the integral cannot be analytically computed in closed form.
 However, as we shall show below, given the Gaussianity of $P(\Delta x|f)$, the moments of the displacement $\Delta x$ can be computed in a straight-forward manner starting from the moments of $f$.
 These moments of the fraction for time spent in the $+$ state further 
 simplify in the long time limit, and hence lead to Eq.~(\ref{eq:Mysels}).

\paragraph*{Scaled cumulant generating function of the fraction of time spent in a state.}
The moment generating function of the time spent in state $+$: $\tau_+$ conditional on being in state $a$ at time $t$ 
is given by
\begin{equation}\label{eq:gen}
G^a_s(t)=\overline{ e^{s\tau_+} \delta(a_t=a)}
\end{equation}
where the overline represents the average with respect of the distribution $P(f)$ 
for a process of duration $t$.
The change in time of the generating function can be found by the methods used~\textit{e.g.}
in~\cite{Gardiner} (Section 7.1) to analyze Birth-Death Master Equations and is given by
\begin{equation}
\frac{d}{dt}
\left(
 \begin{array}{c }
G^+_s	 \\
G^-_s
\end{array}\right)
=\tilde{\TENSOR{M}}_s\left( \begin{array}{c }
G^+_s	 \\
G^-_s
\end{array}
\right)
=\left( \begin{array}{c c }
-k_++s& k_-	 \\
k_+& -k_-	
\end{array}
\right)
\left( \begin{array}{c }
G^+_s	 \\
G^-_s
\end{array}
\right)\,.
\end{equation}
For long times the moment generating function will be dominated by the leading eigenvalue of the matrix $\tilde{\TENSOR{M}}_s$,
which, for the case under consideration, is
\begin{equation}\label{eq:lambda}
\lambda_0(s)=\frac{1}{2} \left(-k_- - k_+ + s + \sqrt{(k_- + k_+ - s)^2 + 4 k_- s}\right)\,.
\end{equation}
The scaled cumulant generating function is defined as (see e.g. \cite{touchette2009large})
$
\lim_{t\to\infty}\frac{1}{t}\log G_s= \lambda_0(s)$
and can be
used to evaluate the long-time expressions of the cumulants of the time spent in a state:
\begin{eqnarray}\label{eq:avet}
\overline{\tau_+}&\simeq&t \frac{\partial}{\partial s} \lambda_0(s)\big|_{s=0}=t\frac{k_-}{k_++k_-}
\\\label{eq:vart}
\overline{ \left(\tau_+-\overline{ \tau_+ }\right)^2}&\simeq&t \frac{\partial^2}{\partial s^2} \lambda_0(s)\big|_{s=0}
=t\frac{2k_-k_+}{(k_++k_-)^3}\\
\overline{\left(\tau_+-\overline{ \tau_+}\right)^3 }&\simeq&t \frac{\partial^3}{\partial s^3} \lambda_0(s)\big|_{s=0}
=-t\frac{6k_-k_+(k_+-k_-)}{(k_++k_-)^5}\label{eq:3t}\\
\overline{\left(\tau_+-\overline{\tau_+} \right)^4 }&\simeq&t\frac{24k_-k_+(k^2_+-k_-k_3-k^2_-)}{(k_++k_-)^7}\\\nonumber
&+&3\left(\overline{ \left(\tau_+-\overline{ \tau_+ }\right)^2}\right)^2\,.
\end{eqnarray}
Higher-order cumulants can also be found from higher derivatives 
of $\lambda$ in (\ref{eq:lambda}).

\paragraph*{Long-time moments of the displacement.}
The above explicit expressions can be used to  evaluate the moments of the displacement. For the average we have
\begin{eqnarray}\label{eq:ave_c_def}
\langle \Delta x\rangle&=&\int\int \Delta x P(\Delta x|f)P(f)  df d\Delta x\\\nonumber
&=&\int \langle \Delta x|f\rangle P(f) df
\end{eqnarray}
where  $\langle \Delta x|f\rangle$
denotes the average displacement for a given $f$ as in eq. (\ref{eq:def_x}). 
Making use of the expression in (\ref{eq:avet})
we obtain:
\begin{eqnarray}\label{eq:ave_c}
\langle \Delta x\rangle&=&Ft\left( \bar{f}\mu_{+}+(1-\bar{f})\mu_{-}\right)\\\nonumber
&=&Ft\left( \frac{k_-}{k_++k_-}\mu_{+}+\frac{k_+}{k_++k_-}\mu_{-}\right)\equiv Ft\bar{\mu}\,.
\end{eqnarray}
Similarly, for the variance one can write: 
\begin{eqnarray}\label{eq:var_c}
&&\left\langle (\Delta x-\langle \Delta x\rangle)^2\right\rangle=\\\nonumber
&=&
\int \left[ 2tD(f)+F^2 (\mu_{+}-\mu_{-})^2(t^2f^2-\bar{f}^2t^2)\right]P(f)  df\\\nonumber
&=&2\left[T\bar{\mu}+F^2(\mu_{+}-\mu_{-})^2\frac{k_-k_+}{(k_-+k_+)^3}\right]t
\end{eqnarray}
which embodies the same result as in (\ref{eq:Mysels}).
In moving from the second line to the third one we have used the expression for the variance of the time spent in $+$ given in (\ref{eq:vart}). As anticipated, the total dispersion involves thermal fluctuations $T\bar{\mu}$  and a term related to the mobility switching which is proportional to the square
of the applied force and the variance of the time spent in a state. 
For the third cumulant one similarly finds
\begin{eqnarray}\nonumber
\left\langle (\Delta x-\langle \Delta x\rangle)^3\right\rangle&=&6TFt(\mu_{+}-\mu_{-})^2\frac{k_-k_+}{(k_++k_-)^3}
\times \\
\label{eq:third-cumulant-of-position}
&&\left[2-\frac{F^2}{T}\frac{(\mu_{+}-\mu_{-})(k_+-k_-)}{(k_++k_-)^2}\right]\,.
\end{eqnarray}
The procedure extends in a simple way to higher moments.\\
The case of several internal states can be treated by  the presented scaled cumulant generating function but requires a slight reformulation that we 
detail in Appendix \ref{appendix:several}. The main difference is that one does not directly compute the scaled cumulant generating function  $\lambda_0(s)$ but evaluates the various moments by means of perturbation theory expanding for small $s$.

\section{Multiple-scales technique}
\label{sec:multiple-scales}
An alternative way 
to compute the first two moments of the displacement 
is to use a multiple-scale technique as 
done, for instance, for rotating Brownian particles in~\cite{aurell2016,marino2016b}.
To this aim it is convenient to rewrite the Fokker-Planck equation 
(\ref{eq:FP})
as 
\begin{eqnarray}\label{eq:fp_mat}
\frac{\partial}{\partial t}
\left(
 \begin{array}{c }
P_+(x)	 \\
 P_-(x)
\end{array}\right)
&=&\left( \begin{array}{c c }
-k_++{\cal L}^\dagger_+& k_-	 \\
k_+& -k_-	+{\cal L}^\dagger_-
\end{array}
\right)
\left( \begin{array}{c }
P_+(x)	 \\
 P_-(x)
\end{array}
\right)
\nonumber\\
&=&
(\TENSOR{K}+{\cal L}^\dagger)\left( \begin{array}{c }
P_+(x)	 \\
 P_-(x)
\end{array}\right)
\end{eqnarray}
where
\begin{eqnarray}\label{eq:L}
{\cal L}^\dagger_a&=&\frac{\partial}{\partial x}\left( -\mu_{a}F+T\mu_{a}\frac{\partial}{\partial x} \right) 
\end{eqnarray}
and $\TENSOR{K}$ is defined in Eq.~\eqref{eq:master}.
In the following we shall make use of the eigendecomposition of $\TENSOR{K}$ which we detail in Appendix~\ref{sec:eigen}.
Since we are seeking a diffusive effective dynamics we will introduce two time scales and adopt the the following scaling
 in space
\begin{subequations}
\label{eq:scales}
\begin{equation}
\label{eq:xalphascales}
\tilde{ x } = \EPS^0  x 
\, , \quad
X  = \EPS^{1}  x 
\, 
\end{equation}
and the following scaling in time
\begin{equation}
\label{eq:tscales}
\theta = \EPS^0 t
\, , \quad
\vartheta = \EPS^{1} t
\, , \quad
\tau = \EPS^{2} t
\, ,
\end{equation}
\end{subequations}
and require that
\begin{equation}
\label{eq:p}
P=P_a(\theta,\vartheta,\tau,\tilde{ x }, X ) \, .
\end{equation}
The first time $\theta$ in (\ref{eq:tscales}) 
is the scale of the jump process, on the order of $k^{-1}$ (jump rates in either direction, provided
they are not widely different). In that time the particle on the average moves a distance about $k^{-1} \mu F$ with scatter on the order $\sqrt{T \mu k^{-1}}$.
These lengths are the first spatial scale in (\ref{eq:xalphascales}). 
The second spatial scale in (\ref{eq:xalphascales})  
is where we observe the motion, and the third time $\tau$ in (\ref{eq:tscales}) is the time scale on which we expect
to see diffusive behavior on that scale. The second time $\vartheta$ in (\ref{eq:tscales}) is the time it takes for the particle to
traverse the large spatial scale moving steadily with the mean velocity $\mu F$; in the present analysis this will be only needed
in an intermediate technical step, and does not by itself generate new physical effects.
With these definitions, $X $ is the the large scale space variable which is of order one only for very large $ x $.
We are interested in finding the effective dynamics in $X$.
The small scale variable $\tilde{ x }$ 
essentially corresponds to the original variables $ x $ 
but is restricted to small scales by imposing periodic
boundary conditions for $P$ over the typical length covered during the relaxation time of $a$. 
As a consequence of the scaling \eqref{eq:scales} and \eqref{eq:p},
the time and spatial derivatives in  turn
into
\begin{eqnarray}
\label{eq:derivatives}
\frac{\partial}{\partial t}
&=& \frac{\partial}{\partial\theta} + \EPS \frac{\partial}{\partial\vartheta} + \EPS^2 \frac{\partial}{\partial\tau}
\, , \\\nonumber
\frac{\partial}{\partial x_i}
&=& \frac{\partial}{\partial\tilde{x}_i} + \EPS \frac{\partial}{\partial X_i}
\, ,
\end{eqnarray}
while the transition matrix
$\TENSOR{K}$ 
remains unchanged.
We  treat $\EPS$ as a small perturbative
parameter and expand $P$ in powers of $\EPS$,
\begin{equation}
\label{eq:pexpansion}
P= P^{(0)} + \EPS P^{(1)} + \EPS^2 P^{(2)} + \ldots
\, ,
\end{equation}
where all $P^{(i)}$ depends  
on the various variables as in eq.~\eqref{eq:p} .
In these variables, $P^{(0)}$ is normalized to one, while all other
$P^{(i)}$ with $i>0$ are normalized to zero.
Plugging \eqref{eq:derivatives} and \eqref{eq:pexpansion} into
\eqref{eq:fp_mat}, and collecting terms of equal powers in
$\EPS$ in the resulting expression, we obtain a
hierarchy of inhomogeneous Fokker-Planck like equations of which
we list the first three (order $\EPS^0$, $\EPS^1$ and $\EPS^2$):
\begin{widetext}
\begin{subequations}
\label{eq:hierarchy}
\begin{eqnarray}
\frac{\partial P^{(0)}}{\partial\theta} - \left( \tilde{\OP{L}}^{\dagger} +  \TENSOR{K}  \right) P^{(0)} & = & 0
\label{eq:hierarchy0}
\\
\frac{\partial P^{(1)}}{\partial\theta} - \left( \tilde{\OP{L}}^{\dagger} +  \TENSOR{K}  \right) P^{(1)} & = &
- \frac{\partial P^{(0)}}{\partial\vartheta}
- \frac{\partial}{\partial X} F\left( \begin{array}{c }
\mu_{+}P^{(0)}_+(x)	 \\
\mu_{-} P^{(0)}_-(x)
\end{array}
\right)
+ 2 T \frac{\partial}{\partial \tilde{x}}\frac{\partial}{\partial X} \left( \begin{array}{c }
\mu_{+}P^{(0)}_+(x)	 \\
\mu_{-} P^{(0)}_-(x)
\end{array}
\right)
\label{eq:hierarchy1}
\\
\frac{\partial P^{(2)}}{\partial\theta} - \left( \tilde{\OP{L}}^{\dagger} +  \TENSOR{K}  \right) P^{(2)} & = &
- \frac{\partial P^{(0)}}{\partial\tau} -\frac{\partial P^{(1)}}{\partial\vartheta}
- \frac{\partial}{\partial X} F\left( \begin{array}{c }
\mu_{+}P^{(1)}_+(x)	 \\
\mu_{-} P^{(1)}_-(x)
\end{array}
\right)\label{eq:hierarchy2}\\\nonumber
&+&T \frac{\partial}{\partial X}\frac{\partial}{\partial X} \left( \begin{array}{c }
\mu_{+}P^{(0)}_+(x)	 \\
\mu_{-} P^{(0)}_-(x)
\end{array}
\right)
+ 2 T \frac{\partial}{\partial \tilde{x}}\frac{\partial}{\partial X} \left( \begin{array}{c }
\mu_{+}P^{(1)}_+(x)	 \\
\mu_{-} P^{(1)}_-(x)
\end{array}
\right)
\end{eqnarray}
\end{subequations}
\end{widetext}

In \eqref{eq:hierarchy}, we introduced the 
tilde over the operators $\tilde{\OP{L}}^\dagger$  to
indicate that it acts on the small scale variables
$\tilde{x}$.
Note that \eqref{eq:hierarchy0} is  the same
equation as \eqref{eq:fp_mat}, with the essential difference, however,
that $P^{(0)}$ obeys periodic boundary conditions in the
variables $\tilde{ x }$.
We follow the standard procedure detailed, e.g., in \cite{pavliotis08}.
We are interested in solutions of \eqref{eq:hierarchy}
which are stationary
on small scales after short-term transients have died out.
Hence, the desired solutions do not depend on $\theta$ such that we
can set $\partial P^{(i)}/\partial\theta = 0$ for all $i$.

\paragraph*{Order $\EPS^0$.}

After relaxation of the fastest timescale,
the periodic solution in $\tilde{x}$  is given by a constant in $\tilde{x}$:  $\rho(\vartheta,\tau,X)$.
For the $a$ variable the system relaxes to the steady state (\ref{eq:ss})
so that we can write:
\begin{equation}
\label{eq:zeroth-order}
P^{(0)}=\VEC{w}\rho(\vartheta,\tau,X)\,.
\end{equation}

\paragraph*{Order $\EPS^1$.}
After relaxation of the fastest time variable the LHS of equation \ref{eq:hierarchy1}
reads: $- \left( \tilde{\OP{L}}^{\dagger} +  \TENSOR{K}  \right) P^{(1)}$.
For a solution to exist we need to require
the RHS to be orthogonal to the left null space of the operator.
Such null space is spanned by constants multiplied by the 
 left null space of $ \TENSOR{K} $ which is given by $\hat{\VEC{w}}=(1,\;1)$
(see Appendix~\ref{sec:eigen}).
 Multiplying the RHS of equation \ref{eq:hierarchy1} corresponds to summing its rows
 and the solvability condition gives:
 \begin{equation}\label{eq:solv1}
0=\frac{\partial \rho}{\partial\vartheta}
+ \frac{\partial}{\partial X} F\left( \mu_{+}\rho w_{+}	+
\mu_{-} \rho w_{-}
\right)=\frac{\partial \rho}{\partial\vartheta}
+ \frac{\partial }{\partial X} \left(F\bar{\mu} \rho\right)
\end{equation}
where 
\begin{equation}\label{eq:mubar}
\bar{\mu}=
w_{+}\mu_{+}+w_{-}\mu_{-}
=\frac{k_-\mu_+}{k_-+k_+}+
\frac{k_+\mu_-}{k_-+k_+}
\end{equation}
 is the same as in \eqref{eq:ave_c} 
since $\overline{f}=w_+$, and where we have used that 
$P^{(0)}$ is independent of $\tilde{x}$.
We can now plug this expression in eq. (\ref{eq:hierarchy1}) which, after the initial relaxation now reads:
\begin{eqnarray} \label{eq:second-order-equation}
 \left( \tilde{\OP{L}}^{\dagger} +  \TENSOR{K}  \right) P^{(1)}  
&=&
 \frac{\partial}{\partial X} \left(F
\left( \begin{array}{c }
\mu_{+}w_{+}	 \\
\mu_{-} w_{-}	\end{array}
\right)\rho\right)\\
&-& \left( \begin{array}{c }
w_{+}	 \\
 w_{-}	\end{array}
 \right)
\frac{\partial }{\partial X} \left(F\bar{\mu} \rho\right)\nonumber\\ \nonumber
&=&
 \frac{\partial}{\partial X} \left(F\left(-\frac{k_+}{k_++k_-}\right)\left(\mu_+-\mu_-\right)\VEC{m}_1\rho\right)
\end{eqnarray}
 where, in moving to the last line we have exploited that the parameters related to the jump process ($k_-$, $k_+$, $w_-$, $w_+$, $\mu_-$, $\mu_+$,
\ldots)
are independent of space and
where 
 \begin{equation}
  \VEC{m}_1=\left( \begin{array}{c }
-1	 \\
 1
\end{array}
\right)\frac{k_-}{k_-+k_+}
  \end{equation}
 is the right eigenvector associated with the non zero eigenvalue of $\TENSOR{K}$, which as required by the solvability condition, is orthogonal to $\hat{\VEC{w}}$ (see Appendix~\ref{sec:eigen}). Since the RHS does not depend on the small spatial scale $\tilde{x}$ we make the ansatz that also $P^{(1)}$ is independent of them. 
Then, $P^{(1)}$ is found by applying $\TENSOR{G}$ the Green's function of the operator $\TENSOR{K}$ given in eq.~\eqref{eq:green} to the RHS. 
The Green's function is defined so that $\TENSOR{K}\TENSOR{G}=-\delta_{i}^j+\VEC{w}\hat{\VEC{w}}$. 
 Making again use of the fact that 
the switching rates (and consequently $\TENSOR{G}$ and $\VEC{w}$)
are independent of space, we can write
\begin{eqnarray}\label{eq:p1}
P^{(1)}&=&
-\frac{\partial\rho}{\partial X} F\TENSOR{G}
\left( \begin{array}{c }
(\mu_{+}-\bar{\mu})w_{+}	 \\
(\mu_{-}-\bar{\mu}) w_{-}	\end{array}
\right)\\\nonumber
&=&
\frac{\partial\rho}{\partial X}F\left(\frac{k_+}{k_++k_-}\right)\left(\mu_+-\mu_-\right)
\TENSOR{G}\VEC{m}_1\\\nonumber
&=&\frac{\partial\rho}{\partial X}F\frac{k_+}{\left(k_++k_-\right)^2}\left(\mu_+-\mu_-\right)\VEC{m}_1
\end{eqnarray}

\paragraph*{Order $\EPS^2$.}
For equation (\ref{eq:hierarchy2}) to admit a solution, we need to require  its RHS to be orthogonal to the left null space of $ \TENSOR{K} $ spanned by $\hat{\VEC{w}}$.
Plugging the solution for $P^{(1)}$ (eq.~\eqref{eq:p1})) and the condition (\ref{eq:solv1}) 
into 
the solvability condition leads to:
\begin{eqnarray}\label{eq:solv2}
 \frac{\partial \rho}{\partial\tau} &=& -F\frac{\partial}{\partial X}\left(\mu_+-\overline{\mu},\, \mu_--\overline{\mu}  \right)P^{(1)}
\\\nonumber
&+&T \frac{\partial}{\partial X}\frac{\partial}{\partial X} \left(\mu_+,\, \mu_- \right)P^{(0)}\\\nonumber
 &=&
  \frac{\partial^2\rho}{\partial X^2}\left(T\overline{\mu}+F^2\frac{k_+k_-}{\left(k_++k_-\right)^3}\left(\mu_+-\mu_-\right)^2 \right)
\end{eqnarray}
where we have made use of the fact that $\left(\mu_+-\overline{\mu},\, \mu_--\overline{\mu}  \right)=-(\mu_+-\mu_-)\frac{k_-}{k_++k_-}\hat{\VEC{n}}_1$.
The connection with the theory of Taylor dispersion can be made explicit by considering the first term on the RHS of the first row of \eqref{eq:solv2} and the explicit
expression of $P^{(1)}$ in \eqref{eq:p1}. The Taylor contribution reads:
\begin{equation}\label{eq:taylor_m}
F^2\left(\mu_+-\overline{\mu},\, \mu_--\overline{\mu}  \right)\TENSOR{G}\left( \begin{array}{c }
(\mu_{+}-\bar{\mu})w_{+}	 \\
(\mu_{-}-\bar{\mu}) w_{-}	\end{array}
\right)
\end{equation}
 which coincides with the known expression (see {\it e. g.} Eq.~(19) in Ref.~\cite{kahlen2017}) if one
 identifies $\mu F$ with the state dependent velocity $u$ in \cite{kahlen2017}.
Composing the two solvability conditions and returning to the original variables we 
have the following SDE for the effective dynamics
\begin{equation}\label{eq:lang_eff}
d x^{eff}=F\bar{\mu}dt+\underbrace{\sqrt{2T\bar{\mu}+2F^2(\mu_{+}-\mu_{-})^2\frac{k_-k_+}{(k_-+k_+)^3}}}_{\sqrt{2D_{eff}}}dW'_t
\end{equation}
showing that the effective equation evolves with average
\begin{equation}\label{eq:ave_m}
\langle \Delta x^{eff}\rangle=\bar{\mu}Ft
\end{equation}
and variance
\begin{eqnarray}\label{eq:var_m}
&&\left\langle\left( \Delta x^{eff}-\langle \Delta x^{eff}\rangle\right)^2\right\rangle=\\\nonumber
&&2\left[T\bar{\mu}+F^2(\mu_{+}-\mu_{-})^2\frac{k_-k_+}{(k_-+k_+)^3}\right]t
\end{eqnarray}
in accordance with Eqs.~\eqref{eq:ave_c} and~\eqref{eq:var_c} 
obtained by using the long time limit of the cumulants 
of the residence time in state $+$.

\section{Generalizations: space-dependent switching rates and several internal states}
\label{sec:generalizations}
In many systems of interest, ({\it e.g.}, molecular motors \cite{watanabe2012mechanical,pietzonka2014fine,kawaguchi2014nonequilibrium}) the transition rates between the internal states may depend on the particle position. 
In general, this may greatly complicate the treatment.
However, under certain assumptions on the spatial dependence, the multiple scale method presented in 
previous section can be applied. For instance, if the switching rates vary appreciably only on the large spatial scales on which the effective diffusion is observed, the operator $\TENSOR{K}$ will depend parametrically on the large spatial coordinate $X$ but will act on the
internal state in the same way.
The zeroth-order equation \eqref{eq:zeroth-order} then still applies, the only
difference being that $\VEC{w}$ is a function of $X$.
The first order solvability equation \eqref{eq:solv1} is also unchanged which means that the drift velocity is
still $F\bar{\mu}$, where $\bar{\mu}$ (which however now depends on $X$) is given in \eqref{eq:mubar}.
The equation to solve on first order \eqref{eq:second-order-equation} is also formally unchanged,
but its coefficients now depend on $X$ and so does the Green's function $G$.

The first point where this type of large spatial scale dependence of the switching rates
enters the analysis is therefore \eqref{eq:p1} which instead will read 
\begin{eqnarray}\label{eq:p1-v2}
P^{(1)}&=&
-\TENSOR{G} \frac{\partial}{\partial X} \left(F
\left( \begin{array}{c }
\mu_{+}w_{+}	 \\
\mu_{-} w_{-}	\end{array}
\right)\rho\right)\\
&+&\TENSOR{G} \left( \begin{array}{c }
w_{+}	 \\
 w_{-}	\end{array}
 \right)
\frac{\partial }{\partial X} \left(F\bar{\mu} \rho\right)\nonumber\,.
\end{eqnarray}
The Green's function can be brought inside the derivative in \eqref{eq:p1-v2}, 
at the price of a correlation term 
proportional $\frac{\partial \TENSOR{G} }{\partial X}$.
The upshot is hence that on the large scale motion is diffusive
with formally the same diffusion coefficient as in \eqref{eq:lang_eff} (now space dependent and following the It\^{o} prescription),
but with a correction to the drift velocity reading:
\begin{equation}
F^2\left(\mu_+-\mu_-\right)\left[\mu_+w_+\frac{\partial}{\partial X}\left(\frac{w_-}{\lambda_1}\right)-
\mu_-w_-\frac{\partial}{\partial X}\left(\frac{w_+}{\lambda_1}\right)\right]\,.
\end{equation}
In principle, a solution with a multiple-scale approach can be found also if the rates depend periodically on space with a period much smaller than the distance travelled by the particle between different jumps in the mobility state. The case in which the rates vary on the same length scale at which the particle moves is more challenging and may not be solvable analytically with a multiple scale approach. 
An extended discussion of a related problem concerning the case of rotational and translational diffusion in periodic potentials can be found in \cite{marino2016dynamics} (see paper IV therein).

The multiple-scale analysis can be straightforwardly generalized to the case in which 
there are more than two discrete states. 
In this Section we extend our formalism to the case of $N$ internal states.
The effective displacement reads as Eq.~\eqref{eq:ave_m} with the average mobility
now displaying the sum over all states: $\overline{\mu}=\sum_i \mu_i w_i$ where $w_i$ represents 
the steady state probability of the discrete process being in state $i$.
Similarly,
the first contribution to the variance, remains proportional to the average mobility as in Eq.~\eqref{eq:var_m}.
The Taylor dispersion contribution to the variance maintains the structure of Eq.~\eqref{eq:taylor_m} and can be written
as
\begin{equation}\label{eq:taylor_n}
F^2\hat{\VEC{w}}\tilde{\TENSOR{\mu}}\TENSOR{G}\tilde{\TENSOR{\mu}}\VEC{w}
\end{equation}
where we have introduced the matrix $\tilde{\TENSOR{\mu}}=\delta_{i}^{j}(\mu_i-\overline{\mu})$ and, as before,
$\TENSOR{G}$ is the Green's function of the transition matrix $\TENSOR{K}$ (on the subspace on non-zero eigenvalues) and $\VEC{w}$ and  $\hat{\VEC{w}}$ are respectively the steady state solution and a unit vector but now extended to   $N$ states.
The main difference is a computational one since explicitly finding the Green's function can become complicated for larger matrices.
An analysis by scaled cumulant
function for several (discrete or continuous) internal states
is carried out in Appendix~\ref{appendix:several} and Appendix~\ref{appendix:translations}.

\section{Coarse-grained properties as a measurement device}
\label{sec:device}
As mentioned in the introduction, the fact that the 
 particle dynamics is influenced by the properties of the 
 friction switching  makes it possible to exploit measurements of the particle position to gain insight about the internal dynamics and evaluate its transition rates.
Remarkably, despite the fact that we are observing the particle on time scales that are longer than the relaxation time of the internal process, we do not only  have access to the equilibrium properties of the internal variable but also to its dynamical ones.
This is possible because we are driving the system away from equilibrium by applying the constant force $F$.
More precisely, from the long-time average displacement of the particle \eqref{eq:ave_c} 
it is possible to measure the average time spent in a state, {\it i. e.}, the equilibrium probability of the internal process $w_+$ and $w_-$ which involve the ratio of the transition rates \eqref{eq:ss}.
To perform such measurement it is necessary to know the friction values corresponding to the two states $\mu_+$ and $\mu_-$ and the 
externally applied force $F$.
The long-term particle dispersion can be exploited to evaluate the magnitude of the friction transition rates.
Expression \eqref{eq:Mysels} 
is the sum of two contributions: a first one being the average of the diffusion constant in the two states (depending on the rate ratio)
and another one
caused by the combination of the external force and the friction fluctuations 
which depends on the variance of the time spent in a given state and involves the magnitude of the transition rates.
The relative importance of the switching dependent contribution depends on how different the two friction coefficients are and on how similar the time scales of diffusion 
and reaction are. Let us introduce the time on which diffusion and drift have comparable effects as
$\tau_D=\overline{\mu} T/(\overline{\mu} F)^2$ and the typical time of the friction switching: $\tau_R=\lambda_1^{-1}=(k_++k_-)^{-1}$.
We then see that
the ratio between the fluctuations due to mobility switching and the thermal ones is proportional to the ratio of the two time scales:
\begin{equation}
\frac{F^2(\mu_{+}-\mu_{-})^2\frac{k_-k_+}{(k_-+k_+)^3}}{T\bar{\mu}}=\frac{\tau_R}{\tau_D}\left(\frac{\mu_{+}-\mu_{-}}{\overline{\mu}}\right)^2w_+w_-\,.
\end{equation}
Such dependence on the time scale can be traced back to the fact that  if the friction dynamics is faster than the diffusive one, many transitions will occur before 
any considerable particle displacement takes place. This
averages out fluctuations in the time spent in a given friction state
 and consequently reduces variance.
 In such case, the particle dynamics will follow Eq.~\eqref{eq:lang} with the state dependent mobility $\mu_a$ replaced by the average one $\overline{\mu}$ in analogy with the cases discussed in \cite{pavliotis08,bo2017} (see \cite{pietzonka2016extreme} for a detailed discussion).
Let us consider a concrete example with
a particle, immersed in water at room temperature, switching between two spherical conformations respectively with radius $\unit[0.6]{\mu m}$ 
 and $\unit[0.4]{\mu m}$ with identical rates $k$, subject to a force of $F=\unit[100]{fN}$.  On average it will move by about $\unit[11]{\mu m\;s^{-1}}$. If the switching occurs at a rate $k=\unit[10]{s^{-1}}$ the position variance will grow as
$\unit[1.4]{\mu m^2\;s^{-1}}$ with the force dependent contribution making up about $36\%$ of the total.

As we have seen one can derive the moments of the particle displacements also for cases with more than two internal states.
In this case, however, knowing the average and variance of the displacement will no longer be sufficient to estimate the transition rates since there are more unknown rates than
known moments. One then has to proceed to measure (and compute) higher moments, a procedure which will cease to be practical as the number of states increases. 
However, for limited systems this may be possible.
Consider for instance, a system composed of three states 
$
+\rightleftharpoons 
0\rightleftharpoons -
$ with four independent rates.
With the presented method,
their estimation then requires four moments. 
Reliably measuring higher moments can be experimentally challenging but moments up to the fourth moment can be obtained. For instance,
the ratio of the fourth to second moment is used to estimate the non Gaussianity of anomalous diffusion (see {\it e.g.} Ref.~\cite{metzler2016non} and references therein).
\section{Conclusions and discussion}
\label{sec:conclusions-discussion}
In this work we have reconsidered a result dating back more than half a century,
that the diffusion of a particle with mobility dependent on an internal state has
two conceptually quite different components. The first term in the effective diffusion is of the standard
thermal type and satisfies an Einstein relation with the effective mobility.
The second term is on the other hand quadratic in the externally applied
force and depends differently on the kinetic rates of transitions between
the internal states. We have pointed out that measuring the first two
moments of displacement hence determines both the rates in both directions when there are
only two internal states and discussed its possible use as an in vitro measurement
device from a modern perspective.
We have generalized these results to more than two internal states, and,
in Appendices, to a continuum of internal states.

The results of this paper
have been derived by a scaled cumulant expansion (large deviations)
and by a multiple-scales calculation.
The extension to the continuous case enabled us to
show that the first and
second order moments of translation of a rotating Brownian particle
can be derived by the scaled cumulant expansion. The continuous internal state is
then the orientation of the particle, an element of the three-dimensional
rotation group SO(3). In a previous contribution~\cite{aurell2016} we
derived this result by a multiple-scales approach.
A conclusion of this work is thus that for the problems considered the multiple-scales
approach can be embedded in the wider mathematical framework of large
deviation theory.

\paragraph*{Acknowledgements.}
We thank Raffaele Marino, Ralf Eichhorn and Erwin Frey for stimulating discussions
 and Udo Seifert and Ramin Golestanian for constructive remarks.
This research was supported by the Academy of Finland through its Center of Excellence COIN (grant no. 251170).
SB wishes to thank Peking University for hospitality.

\appendix

\section{Generalization of the scaled cumulant generating function approach  to several internal states}
\label{appendix:several}

The presence of several internal states requires a slight extension of the scaled cumulant generating function approached presented in Section~\ref{sec:LD}.
The starting point is unchanged,
as we have that the particle evolution at a fixed value of the internal variable is a Gaussian as given in
Eq.~\eqref{eq:p(x|an)}.
The probability of observing a displacement conditioned on a trajectory of the internal variable now depends 
on the fraction of time spent in each state. It is convenient to introduce the column
vector $\VEC{f}$ with entries $f_i$ given by the fraction of time spent in each state $i$ and obviously normalized $\sum_{i=1}^Nf_i=1$.
The expression for the probability is  the same of the one in Eq.~\eqref{eq:p(x|f)} for the two states system as well as Eqs.~\eqref{eq:def_x}:
\begin{eqnarray}\label{eq:p(x|fvec)}
P(\Delta x|\VEC{f};t)
=\frac{
e^{-\frac{\left(\Delta x-\langle \Delta x|\VEC{f}\rangle\right)^2}{4tD(\VEC{f})}}}{\sqrt{4 \pi tD(\VEC{f})}}
\end{eqnarray}
and \eqref{eq:def_d}  with the notable difference that $f$ is replaced by $\VEC{f}$
and that
\begin{eqnarray}
\mu(\VEC{f})=\sum_{i=1}^N \mu_{i}f_i\,.
\end{eqnarray}
The main  difference with the simpler case is that one needs to perform averages over the joint distribution
$P(\VEC{f})=P(f_1,\,f_2,\;\ldots)$.
The average displacement, in analogy with Eq.~\eqref{eq:ave_c_def}, reads:
\begin{eqnarray}\label{eq:ave_c_N}
\langle \Delta x\rangle=Ft\overline{\mu(\VEC{f})}=Ft\sum_{i=1}^N \mu_{i}\overline{f_i}\,.
\end{eqnarray}
where the overline refers to the average over the joint probability $P(\VEC{f})$. In the long-time limit we have that
the average fraction of time spent in a state coincided with the probability of being in that state $\overline{f_i}=w_i$.
For the variance of the displacement one has
\begin{eqnarray}\label{eq:var_c_N}
\left\langle (\Delta x-\langle \Delta x\rangle)^2\right\rangle &=& 2 T t \overline{\mu(\VEC{f})}
+\nonumber \\
  F^2t^2 \int d\VEC{f} &P(\VEC{f})&\sum_{i,\,j} \mu_i\mu_j(f_i-\overline{f_i})(f_j-\overline{f_j})
\end{eqnarray}
requiring the evaluation of the covariance matrix of the fraction of time spent in a state
\begin{equation}\label{sigma}
\Sigma_{i\,j}=\overline{(f_i-\overline{f_i})(f_j-\overline{f_j})}\,.
\end{equation}

To this aim, we can make use of the generating function (in analogy to eq.\eqref{eq:gen})
\begin{equation}\label{eq:gen2}
G^a_{\VEC{s}}=\overline{ e^{t\sum_i s_if_i}\delta(a_t=a)}
\end{equation}
 and exploit the fact that 
 \begin{eqnarray}
t^2\Sigma_{i\,j}=\frac{\partial^2 \log G_{\VEC{s}}}{\partial s_i \partial s_j}\biggr|_{\VEC{s}=\VEC{0}}
\,.
\end{eqnarray}
For long times this can be solved by finding the leading eigenvalue $\lambda_0(\VEC{s})$ of the matrix:
\begin{equation}
\tilde{\TENSOR{M}}_{\VEC{s}}=\TENSOR{K}+\TENSOR{S}
\end{equation}
where we have introduced 
\begin{equation}
\TENSOR{S}=\delta_i^j s_i\,.
\end{equation}
This can be achieved by means 
of perturbation theory
\begin{eqnarray}\label{eq:lambda_pert}
\lambda_0(\VEC{s})&\simeq& \lambda_0(\VEC{0}) + \hat{\VEC{w}}\TENSOR{S}\VEC{w} 
   -\sum_{k>0}\frac{1}{\lambda_k(\VEC{0})}(\hat{\VEC{w}}\TENSOR{S}\VEC{m}_k)(\hat{\VEC{n}}_k\TENSOR{S}\VEC{w})+\ldots\nonumber\\
   &=& 
    \hat{\VEC{w}}\TENSOR{S}\VEC{w} 
   +\hat{\VEC{w}}\TENSOR{S}\TENSOR{G}\TENSOR{S}\VEC{w}+\ldots
\end{eqnarray}
where $\lambda_k(\VEC{0})$ are the eigenvalues of the unperturbed matrix $\TENSOR{K}$ and $\VEC{m}_k$, $\hat{\VEC{n}}_k$ respectively its right and left eigenvectors. For consistence with earlier notation we have denoted the eigenvectors of the zero eigenvector as $\VEC{m}_0=\VEC{w}$, $\hat{\VEC{n}}_0=\hat{\VEC{w}}$.
We have also introduced the Green's function $-\sum_{k>0}\frac{1}{\lambda_k(\VEC{0})}\VEC{m}_k\hat{\VEC{n}}_k=\TENSOR{G}$.
The covariance then reads
 \begin{equation}
t^2 \Sigma_{i\,j}=\left(\hat{w}_iG_{i\,j}w_j+\hat{w}_jG_{j\,i}w_i\right)t
 \end{equation}
 where no summation is to be taken over the repeated indices.
 With this  expression the term proportional to the force squared in
 \eqref{eq:var_c_N}, reads
  \begin{eqnarray}
 tF^2\sum_{i,\,j}\mu_i \Sigma_{i\,j}\mu_j&=&tF^2\mu_i(\hat{w}_iG_{i\,j}w_j+\hat{w}_jG_{j\,i}w_i)\mu_j\nonumber\\
&=& 2tF^2\hat{\VEC{w}}\tilde{\TENSOR{\mu}}\TENSOR{G}\tilde{\TENSOR{\mu}}\VEC{w}
 \end{eqnarray}
 corresponding with the diffusion related to the term in Eq.~\eqref{eq:taylor_n},
where we have exploited the biorthogonality $\overline{\mu}^2\hat{\VEC{w}}\TENSOR{G}\VEC{w}=0$. 

\section{Eigen decomposition of $\TENSOR{K}$ for two states}
\label{sec:eigen}
In this section we will present the eigen-decomposition of the matrix $\TENSOR{K}$. 
Since it generates a Markov process, the largest eigenvalue of
$\TENSOR{K}$ is $\lambda_0=0$ and its associate right eigenvector
 gives the steady state distribution
 \begin{equation}
\VEC{w}=\left( \begin{array}{c }
w_{+}	 \\
 w_{-}
\end{array}
\right)=
\left( \begin{array}{c }
\frac{k_-}{k_-+k_+}	 \\
\frac{k_+}{k_-+k_+}
\end{array}
\right)
\end{equation}
 and its left eigenvector 
 \begin{equation}
 \hat{\VEC{w}}=(1,\,1)
 \end{equation} ensures probability conservation. 
 With
 \begin{equation}
 \TENSOR{K}\VEC{w}=0\qquad\qquad \hat{\VEC{w}} \TENSOR{K}=0
 \end{equation}
and
 \begin{equation}
 \hat{\VEC{w}}\VEC{w}=1\,.
  \end{equation}
  The second eigenvalue is given by
  \begin{equation}
\lambda_1=-\left(k_-+k_+\right)
  \end{equation}
  and its eigenvectors are
  \begin{equation}\label{eq:e_vec1}
  \VEC{m}_1=\left( \begin{array}{c }
-1	 \\
 1
\end{array}
\right)\frac{k_-}{k_-+k_+}\qquad   \hat{\VEC{n}}_1=\left(-\frac{k_+}{k_-},\,1\right)
  \end{equation}
  where
   \begin{equation}
 \TENSOR{K}\VEC{m}_1=\lambda_1\VEC{m}_1\qquad\qquad \hat{\VEC{n}}_1 \TENSOR{K}=\lambda_1 \hat{\VEC{n}}_1
 \end{equation}
 and
 \begin{equation}
 \hat{\VEC{n}}_1\VEC{m}_1=1\,.
  \end{equation}
This provides a biorthonormal decomposition with:
\begin{equation}
 \hat{\VEC{n}}_1\VEC{w}=0 \qquad \hat{\VEC{w}}\VEC{m}_1=0
\end{equation}
and the completeness relation
\begin{equation}
\VEC{w} \hat{\VEC{w}}+ \VEC{m}_1\hat{\VEC{n}}_1=\left( \begin{array}{c c }
1&0	 \\
0&1
\end{array}
\right)
\end{equation}
The Green's function
\begin{equation}\label{eq:green}
\TENSOR{G}=-\frac{1}{\lambda_1}\VEC{m}_1\hat{\VEC{n}}_1 = \frac{1}{\left(k_-+k_+\right)^2}\left(\begin{array}{cc} k_{+} & -k_{-} \\ -k_{+} & k_{-} \end{array}\right) 
\end{equation}
ensures that for vectors orthogonal to the left null space of $\TENSOR{K}$ (parallel to $\VEC{m}_1$)
\begin{equation}
-\TENSOR{K}\TENSOR{G}\VEC{m}_1=\frac{1}{\lambda_1}\TENSOR{K}\VEC{m}_1\hat{\VEC{n}}_1\VEC{m}_1=
\VEC{m}_1\hat{\VEC{n}}_1\VEC{m}_1=\VEC{m}_1
\end{equation}

\section{Large deviation analysis with internal continuous degree of freedom}
\label{appendix:translations}

In this Appendix we make the connection to our earlier work 
on coarse-graining of diffusion in space and over orientations~\cite{aurell2016,marino2016b}.
The starting point is the same as in the preceding discussion, except that
the  internal variable $\alpha$
representing the internal state now lives
in a compact manifold ${\cal M}$ with volume element $\sqrt{g}$. 
We also assume that the spatial motion of the particle takes place in $D$-dimensional space.
Rotations in $D=2$ can hence be represented by an internal coordinate on a circle 
(Lie group $SO(2)$), rotations in 
$D=3$ by an internal coordinate in the solid upper hemisphere (Lie group $SO(3)$), and so on.
In~\cite{marino2016b} the case of motion in two-dimensional complex space ($D=4$) and 
the group $SU(2)$ was considered as well.
Whatever the internal state the process is described by
\begin{equation}\label{eq:overdampedequations}
\text{d}x_{i}= v_i(\alpha)\text{d}t+\sqrt{2T}\mathbf{\mu}^{\frac{1}{2}}_{ij} (\alpha)\text{d}W_j
\end{equation}
As in the previous discussion $\mathbf{\mu}(\alpha)$ in \eqref{eq:overdampedequations} is a mobility matrix and
$\mathbf{\mu}^{\frac{1}{2}}$ is its matrix square root,
$\text{d}W_j$ is the standard Wiener increment, $v_i(\alpha)=\mathbf{\mu}(\alpha)_{ij}F_i$ is a drift velocity and 
$F$ is an applied force, assumed constant in space and time.
We assume that $\mathbf{\mu}(\alpha)$ depends smoothly on $\alpha$.

Now suppose that $\alpha$ obeys a dynamic law independent of spatial position $x$
such that in a total time $t$ the coordinate $\alpha$ can be found in a set $A\subset {\cal M}$
for a time $t_A$. In parallel with the discussion in the main text
we can introduce $f_A(t)=\frac{t_A}{t}$, the fraction of the time 
$\alpha$ can be found in $A$ in a total time $t$. Suppose further that 
${\cal M}$ is divided up in a collection of small sets $A_k$ with zero-measure intersections
and which together cover ${\cal M}$. Approximately the motion must then be
same as with a finite number of internal states discussed in the preceding
Section~\ref{appendix:several}.

In the limit that all the sets $A_k$ are small
we assume that 
there is a normalized empiric measure $m(\alpha)\sqrt{g}$ smoothly related to the volume element
such that $f_{A_k} = \int_{A_k}\sqrt{g} m(\alpha)$ and $\int_{\cal M}\sqrt{g} m(\alpha) = 1$. 
In analogy with the previous case this leads to an overall
 mobility dependent on the empiric measure
\begin{equation}\label{eq:overall-mobility-smooth}
  \mathbf{\mu}[m]
  = \int_{\cal M} \sqrt{g}\, m(\alpha) \mathbf{\mu}(\alpha)
\end{equation}
and an overall drift velocity
\begin{equation}\label{eq:overall-drift-smooth}
\VEC{v}[m]_i = \int_{\cal M} \sqrt{g}\, m(\alpha) \mathbf{\mu}(\alpha)_{ij} F_j
\end{equation}
In both \eqref{eq:overall-mobility-smooth} and \eqref{eq:overall-drift-smooth} 
the integrals are performed with  the
empirical measure, and  $\mathbf{\mu}[m]$ and $\VEC{v}[m]_i$
are hence both linear functionals
of the scalar function $m(\alpha)$.

We now further postulate a functional $Q[m(\alpha),t]$ such that ${\cal D}m Q[m]$ is
the probability of an empirical measure
in a functional volume element ${\cal D}m$ around $m(\alpha)\sqrt{g}$. 
$Q$ could be taken the limit 
of finite-dimensional probability distributions
$Q[\{f_{V_k}\},t]$
where $\{f_{V_k}\}$ is a finite set of empirical frequencies analogous to $\VEC{f}$ in the previous section.
We define the average of the empirical measures as
\begin{eqnarray}
  \overline{m(\alpha)} &=& \int {\cal D}m Q[m] m(\alpha)\\
  \overline{m(\alpha) m(\alpha')} &=& \int {\cal D}m Q[m] m(\alpha)m(\alpha')\\
  &\vdots& \nonumber
\end{eqnarray}
The effective velocity will then be the average with respect to $Q$ of the drift velocity 
in (\ref{eq:overall-drift-smooth}) \textit{i.e.}
\begin{equation}\label{eq:v-eff-def}
\VEC{v}^{eff}_i = \int {\cal D}m Q[m] \VEC{v}[m]_i = \int_{\cal M} \sqrt{g}\, \overline{m(\alpha)} \mathbf{\mu}(\alpha)_{ij} F_j
\end{equation}
In the long time limit $\overline{m(\alpha)}$ will be the same as the steady
state distribution $\rho^{ss}(\alpha)$, as will be verified below.  

Introducing for compactness $\delta \VEC{X}=\Delta \VEC{X} - t \VEC{v}^{eff}$ we can
compute the centered second moment
\begin{eqnarray}
  \label{eq:centered-secondmoment}
  \left< \delta X_i \delta X_j \right> &=& 2 T t \int {\cal D}m Q[m] \mathbf{\mu}_{ij}[m] +\nonumber \\
  t^2 \int {\cal D}m &Q[m]& (v_i[m]-v_i^{eff})(v_j[m]-v_j^{eff})
\end{eqnarray}

The two functional integrals in (\ref{eq:centered-secondmoment})
can respectively be written
\begin{eqnarray}
  \label{eq:secondmoment-detailed-1}
  \int {\cal D}m Q[m] \mathbf{\mu}_{ij}[m] &=& \int_{\cal M}\sqrt{g}\, \overline{m(\alpha)} \mathbf{\mu}_{ij}(\alpha) =  \mathbf{\mu}_{ij}^{eff}
\end{eqnarray}
and
\begin{widetext}
\begin{eqnarray}
  \label{eq:secondmoment-detailed-2}
  \int {\cal D}m Q[m] (v_i[m]-v_i^{eff})(v_j[m]-v_j^{eff}) &=& \int {\cal D}m Q[m] \int_{\cal M}\sqrt{g}\, \int_{\cal M'}\sqrt{g'}\,
  \left(m(\alpha)-\overline{m(\alpha)}\right) \mathbf{\mu}_{ik}(\alpha)
  \left(m(\alpha')-\overline{m(\alpha')}\right) \mathbf{\mu}_{jl}(\alpha') F_k F_l \nonumber \\
  &=& \int_{{\cal M}\, {\cal M'}}\sqrt{g}\sqrt{g'}\, \left(\overline{m(\alpha)m(\alpha')}-\overline{m(\alpha)}\,\overline{m(\alpha')}\right)
  \mathbf{\mu}_{ik}(\alpha) \mathbf{\mu}_{jl}(\alpha') F_k F_l\,.
\end{eqnarray}
\end{widetext}
We introduce the moment generating functional of the time spent at internal position $\alpha$, conditional on ending
up at internal position $\alpha'$ at time $t$. This moment generating functional depends on a function $s(\alpha)$ and is defined as
\begin{equation}
    G^{\alpha'}[s] = \left< e^{\int_0^t s(\alpha_t) dt}\delta(\alpha_t-\alpha')\right>
\end{equation}
We introduce also the probability $Q[m]^{\alpha'}$ of observing a frequency distribution
in a functional volume element around measure $m(\alpha)$ for a process ending at $\alpha'$ at time $t$.
Then
\begin{equation}
  G^{\alpha'}[s] = \int {\cal D}m Q[m]^{\alpha'}\, e^{t \int_G\sqrt{g}\, s(\alpha) m(\alpha)}
\end{equation}
We now assume that the time development of $\alpha$ is a stochastic process with generator ${\cal L}$.
The moment generating functional then obeys 
\begin{equation}
    \partial_t G^{\alpha'}[s] = {\cal L}^{\dagger} G[s]^{\alpha'} + s(\alpha') G^{\alpha'}[s]
\end{equation}
where ${\cal L}^{\dagger}$ is the adjoint of the generator (Fokker-Planck operator for a diffusion process) acting at internal coordinate $\alpha'$.
For long times the moment generating function will be almost independent of $\alpha'$ and 
dominated by
 $ t \lambda_0[s]$, where $ \lambda_0[s]$ is
the largest eigenvalue of the operator ${\cal L}^{\dagger} + s(\alpha)$.
If $s=0$ this largest eigenvalue must be zero, corresponding to the stationary probability distribution for the
stochastic process in $\alpha$. Since ${\cal M}$ is assumed compact this eigenvalue must further be separated by a gap from all
the decaying states, and $\lambda_0[s]$ must therefore be close to zero for $s$ sufficiently small.

To evaluate the terms in (\ref{eq:secondmoment-detailed-1}) and (\ref{eq:secondmoment-detailed-2})
we consider as before in the long term limit of the generating functional and its functional logarithmic derivatives evaluated at $s=0$:
\begin{eqnarray}
  \frac{\delta \log G[s]}{\delta s(\alpha)}|_{s=0}  &=& t \overline{m(\alpha)}\sqrt{g} \nonumber \\
  \frac{\delta^2 \log G[s]}{\delta s(\alpha) \delta s(\alpha') }|_{s=0} &=&
  t^2 \left(\overline{m(\alpha) m(\alpha')}-\overline{m(\alpha)}\, \overline{m(\alpha')}\right)
  \sqrt{g}\sqrt{g'} \nonumber
\end{eqnarray}
These functional derivatives are in the long time limit
as $t\frac{\delta \lambda[s]}{\delta s(\alpha)}$
and $t\frac{\delta^2 \lambda[s]}{\delta s(\alpha) \delta s(\alpha') }$
and the eigenvalue can be computed in 
in time-independent perturbation theory
\begin{widetext}
\begin{eqnarray}
\label{eq:Rayleigh-Schrodinger-second-order}
 \lambda_0[s] &=& \lambda_0[0] + \int_{\cal M} \sqrt{g} n_0(\alpha) s(\alpha) m_0(\alpha)\\\nonumber
  && \qquad -   \sum_{k>0}\frac{1}{\lambda_k[0]}\int_{\cal M}\sqrt{g} n_0(\alpha)s(\alpha)m_k(\alpha) \int_{{\cal M}'}\sqrt{g'} n_k(\alpha')s(\alpha')m_0(\alpha')+\ldots
\end{eqnarray}
\end{widetext}
where $m_k$ and $n_k$ are the right and left eigenfunctions of ${\cal L}^{\dagger}$ corresponding to eigenvalue $\lambda_k[0]$
and the scalar product is $(f,g)=\int_{\cal M}\sqrt{g}\, fg$.
The average of the empirical measure is therefore (in the long-time limit) 
the first functional derivative of (\ref{eq:Rayleigh-Schrodinger-second-order}) with respect to $s(\alpha)$,
which gives 
\begin{eqnarray}
  \label{eq:first-moment-m}
  \overline{m(\alpha)} &=& n_0(\alpha) m_0(\alpha)
\end{eqnarray}
The left eigenfunction $n_0$ is the zero mode of the generator ${\cal L}$. For diffusions, where all the derivatives in ${\cal L}$ are on the right, $n_0$ must then be a constant. By orthonormality $(n_0,m_0)=1$ and it is convenient to  take $n_0=1$ and $m_0$ a normalized probability distribution on ${\cal M}$ \textit{i.e.} $\int_{\cal M}\sqrt{g}\, m_0 =1$. 
This normalized probability distribution is at the same time the right
eigenvector of the adjoint of the generator
with eigenvalue zero \textit{i.e.} 
the stationary probability distribution $\rho^{ss}(\alpha)$
of the stochastic process, as already observed above.
For the first term in (\ref{eq:centered-secondmoment}) we combine (\ref{eq:secondmoment-detailed-1}) and (\ref{eq:first-moment-m}) and find a first term in an effective diffusion tensor
\begin{equation}
  \label{eq:effective-diffusion-1}
  D^{(1)}_{ij}          = T \mathbf{\mu}_{ij}^{eff}
\end{equation}
where the effective mobility tensor has been 
introduced in \eqref{eq:secondmoment-detailed-1}

The covariance of the empirical measure is
  \begin{eqnarray}
        \label{eq:second-moment-m}
 && \left(\overline{m(\alpha) m(\alpha')}-\overline{m(\alpha)}\, \overline{m(\alpha')}\right)
  = \\\nonumber
  &&\frac{1}{t} {\cal G}(\alpha,\alpha') m_0(\alpha') +
  \left(\hbox{$\alpha\leftrightarrow \alpha'$} \right) 
\end{eqnarray}
where we have introduced a Green's function
\begin{equation}
\label{eq:Greens-function-def}
{\cal G}(\alpha,\alpha') = -\sum_{k>0}\frac{1}{\lambda_k[0]} m_k(\alpha) n_k(\alpha')
\end{equation} 
and used that $n_0(\alpha)$ is a constant that we can set to one.
The Green's function satisfies the equation
\begin{equation}
{\cal L}^{\dagger}{\cal G}(\alpha,\alpha') = - \delta(\alpha-\alpha') + \Pi_0
\end{equation}
where $\Pi_0$ is a projection operator that acts on a function $f$ as $[\Pi_0 f](\alpha) = (n_0,f) m_0(\alpha)$.

The second term in (\ref{eq:centered-secondmoment}) \textit{i.e.} 
the contribution from the functional integral in (\ref{eq:secondmoment-detailed-2})
involves two mobility matrices $\mu_{ik}(\alpha)$ and $\mu_{jl}(\alpha')$ and
the Green's function ${\cal G}(\alpha,\alpha')$.     
A convenient representation is to first 
introduce an auxiliary function
\begin{eqnarray}
  \label{eq:lambda-def}
  \lambda_{jl}(\alpha) &=& \int_{\cal M'}\sqrt{g'}\,   m_0(\alpha') {\cal G}(\alpha,\alpha') \mathbf{\mu}_{jl}(\alpha')  
\end{eqnarray}
Acting with ${\cal L}^{\dagger}$ on $\lambda$ gives
\begin{eqnarray}
    \label{eq:solvability}
          {\cal L}^{\dagger} \lambda_{jl}(\alpha) &=& - \mathbf{\mu}_{jl}(\alpha)m_0(\alpha)
          + \Pi_0[\mathbf{\mu}_{jl}m_0](\alpha) \nonumber \\
          &=& - \left(\mathbf{\mu}_{jl}(\alpha)-\mathbf{\mu}^{eff}_{jl}\right) m_0(\alpha)  
\end{eqnarray}
We observe that
\begin{equation}
  \label{eq:solvability-3}
  \int_{\cal M}\sqrt{g} \lambda_{jl}(\alpha) = \int_{\cal M'}\sqrt{g'}\, m_0(\alpha') \mathbf{\mu}_{jl}(\alpha')
                                              \int_{\cal M} \sqrt{g} {\cal G}(\alpha,\alpha') \nonumber 
\end{equation}
where the inner integral is
\begin{equation}
  \label{eq:solvability-4}
     \int_{\cal M} \sqrt{g} {\cal G}(\alpha,\alpha')= - \sum_{k>0}\frac{1}{\lambda_k[0]} n_k(\alpha')  \int_{\cal M} \sqrt{g} m_k(\alpha) = 0  
\end{equation}
and the last equality follows because all the right eigenfunctions $m_k$ with $k>0$ must be orthogonal to $n_0$ which is a constant.
The second term can thus be written
\begin{equation}
  \label{eq:solvability-5}
  D^{(2)}_{ij} = A_{ijkl}F_kF_l
\end{equation}
where the fourth-order tensor $A$ is
\begin{equation}
\label{eq:solvability-6}
  A_{ijkl} = \int_{\cal M}\sqrt{g} \left(\mu_{ik}(\alpha) - \mu_{ik}^{eff}\right)\lambda_{jl}(\alpha) + \hbox{perm.} 
\end{equation}
For the cases studied in~\cite{aurell2016,marino2016b} 
$\mathbf{\mu}_{jl}(\alpha)$ equals $[R(\alpha)\mu^B R^{-1}(\alpha)]_{jl}$
where $R(\alpha)$ is a rotation matrix and $\mu^B$ is a mobility matrix in the body frame,
and $m_0$ is the uniform (Haar) measure. As discussed therein
$\mathbf{\mu}^{eff}_{jl}$ then equals $\frac{1}{3}\hbox{Tr}[\mathbf{\mu}] \mathbf{1}_{jl}$
and the quantity in the parenthesis on the right-hand side of \eqref{eq:solvability} is the traceless part of the
mobility matrix. In~\cite{aurell2016} and~\cite{marino2016b} this quantity was denoted $\tilde{\gamma}^{-1}$.  
Eqs.~\eqref{eq:solvability}, \eqref{eq:solvability-5} and \eqref{eq:solvability-6} 
generalize the corresponding expressions from
\cite{aurell2016} to cases where steady state probability is not necessarily uniform.

\section{Higher moments for continuous internal variables}
\label{appendix:higher-moments}
In this Appendix we compute the centered third moment with the scaled cumulant expansion technique
and show that it is linear in time $t$. It can be written as a sum of 
three terms given below in \eqref{eq:third-order-first-term} and \eqref{eq:third-centered-moment-2}. 
The starting point is 
\begin{widetext}
\begin{eqnarray}
  \left< \delta X_i \delta X_j \delta X_k \right> &=& 2 T t^2 \int {\cal D}m Q[m] \mathbf{\mu}_{ij}[m](v_k[m]-v_k^{eff}) +\hbox{perm.} 
\nonumber \\
\label{eq:third-centered-moment}
&& \qquad + t^3 \int {\cal D}m Q[m] (v_i[m]-v_i^{eff})(v_j[m]-v_j^{eff})(v_k[m]-v_k^{eff})
\end{eqnarray}
where the second term generalizes \eqref{eq:second-moment-m} and \eqref{eq:secondmoment-detailed-2} and can be written
\begin{equation}
\label{eq:third-centered-moment-2}
\hbox{$2^{nd}$}= \int_{\cal M} \int_{\cal M'} \int_{\cal M''}
  \left[\sqrt{g}\sqrt{g'}\sqrt{g''} t^3 \overline{\left(m(\alpha)-\overline{m(\alpha)}\right)\left(m(\alpha')-\overline{m(\alpha')}\right)\left(m(\alpha'')-\overline{m(\alpha'')}\right)} \right]
  \mu_{il}(\alpha) \mu_{jm}(\alpha') \mu_{kn}(\alpha'') F_l F_m F_n 
\end{equation}
\end{widetext}
Comparing to (\ref{eq:secondmoment-detailed-2})
\eqref{eq:solvability-5} and \eqref{eq:solvability-6} one sees that
the first term in (\ref{eq:third-centered-moment}) is linear in time and given by
\begin{equation}
\label{eq:third-order-first-term}
  \left< \delta X_i \delta X_j \delta X_k \right>^{(1)} = t 2T \left(A_{ijkl}F_l +  \hbox{perm.}\right) 
\end{equation}
The tensor $A$ is given in \eqref{eq:solvability-6} above.
Turning to the other term, the expression inside the inner square bracket of \eqref{eq:third-centered-moment-2}
is
$\frac{ \delta^3 \log G[s]}
{\delta s(\alpha) \delta s(\alpha') \delta s(\alpha'')}$
which is equal to 
$t\frac{\delta^3 \lambda_0[s]}
{\delta s(\alpha) \delta s(\alpha') \delta s(\alpha'')}$.
Also this term is therefore linear in time.
To compute it we thus need the third order perturbation of an energy which is
\begin{widetext}
  \begin{eqnarray}
\label{eq:third-order}
    \lambda_0^{(3)}[s] &=& \sum_{m>0 }\sum_{n>0}\frac{<0|s|m><m|s|n><n|s|0>}{\lambda_n[0] \lambda_m[0]}  - <0|s|0> \sum_n \frac{<0|s|n><n|s|0>}{(\lambda_n[0])^2}\nonumber \\
    &=& \int_{\cal M}\sqrt{g} \int_{\cal M'}\sqrt{g'} \int_{\cal M''}\sqrt{g''} \, n_0(\alpha) s(\alpha) {\cal G}(\alpha,\alpha') s(\alpha') {\cal G}(\alpha',\alpha'') s(\alpha'') m_0(\alpha'') \nonumber \\
    && - \int_{\cal M}\sqrt{g}\, n_0(\alpha) s(\alpha) m_0(\alpha) \times \int_{\cal M'}\sqrt{g'} \int_{\cal M''}\sqrt{g''}\, 
         n_0(\alpha') s(\alpha') {\cal G}^{(2)}(\alpha',\alpha'') s(\alpha'')m_0(\alpha'')
\end{eqnarray}
\end{widetext}
where  ${\cal G}^{(2)}(\alpha',\alpha'')=\sum_{n>0}m_n(\alpha')n_n(\alpha'')\frac{1}{(\lambda_n[0])^2}$
satisfies $\left({\cal L}^{\dagger}\right)^2 {\cal G}^{(2)}(\alpha',\alpha'') = \delta(\alpha'-\alpha'') - \Pi_0$. This means that the the second term in \eqref{eq:third-centered-moment} can be written
as a sum of two terms
\begin{equation}
  \label{eq:third-centered-moment-2} 
 \left< \delta X_i \delta X_j \delta X_k \right>^{(2)} = t \left(B^{(1)}_{ijklmn} -B^{(2)}_{ijklmn}\right)   F_lF_m F_n 
\end{equation}
where the two sixth order tensors are given by
\begin{widetext}
\begin{equation}
\label{eq:B-1}
B^{(1)}_{ijklmn} =
\int_{\cal M}\sqrt{g} \int_{\cal M'}\sqrt{g'} \int_{\cal M''}\sqrt{g''} \, n_0(\alpha) {\cal G}(\alpha,\alpha') {\cal G}(\alpha',\alpha'') m_0(\alpha'') \mu_{il}(\alpha)
\mu_{jm}(\alpha') \mu_{kn}(\alpha'') + \hbox{perm.}
\end{equation}
and
\begin{equation} 
 \label{eq:B-2}
B^{(2)}_{ijklmn} =
\int_{\cal M}\sqrt{g} n_0(\alpha) m_0(\alpha) \mu_{il}(\alpha)\cdot \int_{\cal M'}\sqrt{g'} \int_{\cal M''}\sqrt{g''} \, 
n_0(\alpha') {\cal G}^{(2)}(\alpha',\alpha'') m_0(\alpha'') \mu_{jm}(\alpha') \mu_{kn}(\alpha'') + \hbox{perm.}
\end{equation}
\end{widetext}
For the term $B^{(1)}$ we can use the same auxiliary quantity as above 
\begin{equation}
\lambda_{kn}(\alpha') = \int_{\cal M''} \sqrt{g''}  {\cal G}(\alpha',\alpha'') m_0(\alpha'') \mu_{kn}(\alpha'')
\end{equation}
and further another auxiliary quantity
\begin{equation}
\lambda^{L}_{il}(\alpha') = \int_{\cal M} \sqrt{g}  n_0(\alpha) {\cal G}(\alpha,\alpha')  \mu_{il}(\alpha)
\end{equation}
which satisfies $({\cal L}_0 \lambda^{L}_{il})(\alpha') = \mu_{il}(\alpha') - \mu_{il}^{eff}$
and which together give
\begin{equation}
 \label{eq:B-1-2}
B^{(1)}_{ijklmn} = \int_{\cal M'}\sqrt{g'} \lambda^L_{ij}(\alpha') \mu_{jm}(\alpha') \lambda_{kn}(\alpha') + \hbox{perm.}
\end{equation}
For the term $B^{(2)}$ we introduce yet another auxiliary quantity
\begin{equation}
\lambda^{(2)}_{kn}(\alpha') = \int_{\cal M''} \sqrt{g''}  {\cal G}^{(2)}(\alpha',\alpha'') m_0(\alpha'') \mu_{kn}(\alpha'')
\end{equation}
in terms of which we have 
\begin{equation}
 \label{eq:B-2-2}
B^{(2)}_{ijklmn} =\mu_{il}^{eff} \int_{\cal M'}\sqrt{g'} \mu_{jm}(\alpha') \lambda^{(2)}_{kn}(\alpha') + \hbox{perm.}
\end{equation}
When $m_0$ is the Haar measure the averages in 
 \eqref{eq:B-1-2}
 \eqref{eq:B-2-2}
can be computed by invariance arguments as in~\cite{aurell2016} and~\cite{marino2016b}.

\end{document}